\documentclass[12pt]{iopart}

\usepackage{iopams}

\begin{document}

\title{Collective Dipole Model of AdS/CFT and Higher Spin Gravity}

\author{Antal Jevicki, Kewang Jin and Qibin Ye}

\address{Department of Physics, Brown University, Providence, RI 02912, USA}

\eads{\mailto{antal\_jevicki@brown.edu}, \mailto{kewang\_jin@brown.edu}, \mailto{qibin\_ye@brown.edu}}

\begin{abstract}

We formulate a first quantized construction of the AdS$_{d+1}$/CFT$_d$ correspondence using the bi-local representation of the free $d$-dimensional large $N$ vector model. The earlier reconstruction of AdS$_4$ higher-spin gravity provides a scheme where the AdS spacetime (and higher-spin fields) are given by the composite bi-local fields. The underlying first quantized, world-sheet picture is extracted in the present work and generalized to any dimension. A higher-spin AdS particle model is shown to emerge from the collective bi-particle system of Minkowski particles through a canonical transformation. As such this construction provides a simple explicit mechanism of the AdS/CFT correspondence.

\end{abstract}

\maketitle

\section{Introduction}

The AdS/CFT correspondence \cite{Maldacena:1997re} represents a major tool in our understanding of non-perturbative phenomena in gauge theory (and other related systems). Insight into the mechanism behind this duality was obtained through various different tools, such as large $N$ expansion, D-branes and higher symmetries. The explicit construction, even in the simplest models, has not been achieved yet (except in special limits or sub-sectors of the full theory such as the 1/2 BPS case). What characterizes the correspondence is the emergence of AdS spacetime (and of extra Kaluza-Klein dimensions) and even more remarkably of gravitational and stringy degrees of freedom. 

Recently a very simple model (the $O(N)$ vector model) has been studied with its duality \cite{Klebanov:2002ja, Sundborg:2000wp} to AdS higher-spin gravity of Vasiliev \cite{Vasiliev:1995dn, Sorokin:2004ie}. This proposal which identifies the critical points of the 3d $O(N)$ vector model with two versions of the 4d Vasiliev theory has received solid support \cite{Giombi:2009wh} and definite degree of understanding. Interesting studies of the AdS$_3$/CFT$_2$ correspondence are also done in \cite{Henneaux:2010xg}. One particular approach that we have pursued is the construction of the (dual) AdS theory from the CFT in terms of collective fields \cite{Das:2003vw, Koch:2010cy}. Collective fields extend the space of (conformal) operators (and conserved currents) that are usually used for ``holographic'' comparisons of correlators and amplitudes at the boundary. They provide a framework for reconstruction of the AdS theory in the bulk and with interactions. In the specific case of the $O(N)$ vector models these fields are given by bilocal invariants representing scalar products of basic local fields. It was demonstrated in \cite{Koch:2010cy} in the example of 3d free CFT that the bilocal field contains fully the additional (radial) AdS dimension and also the infinite sequence of fields with growing spins. This construction (done in the light-cone gauge) provides a full one-to-one map between (field) observables of the field theory and fields of the higher-spin gravity. Other versions which involve use of renormalization group schemes were given in \cite{Lee:2009ij, Douglas:2010rc}. An effective action approach was also developed in \cite{Bekaert:2010ky} through coupling of higher spin fields to conformal scalar field theory. The correspondence between various approaches remains to be studied.

The reconstruction of AdS spacetime from the bilocal spacetime turned out to be given by a canonical transformation of phase space (rather than just a coordinate transformation). In the particular case of AdS$_4$/CFT$_3$, the transformation can be essentially seen as given by a simple Fourier transformation. With the field map in place one might ask the question of the first quantized (world-sheet) picture behind this correspondence. In the ${\cal N}=4$ SYM case the world-sheet picture given by the integrable Yang-Baxter system played a central role and it is of interest to illuminate it in the present case of $O(N)$ theories. Instead of the string we now have a discrete bi-particle system which we call {\it the collective dipole}. In what follows we describe and study its classical dynamics and work out the details of its map to AdS. This we do in any dimension $d$ showing the reconstruction of higher-spin system in AdS through a canonical transform. As such the collective dipole offers possibly the simplest system for a deeper understanding of the emergence of extra dimension and higher-spin degrees of freedom. A dipole picture was originally contemplated by Flato and Fronsdal \cite{Flato:1978qz} and subsequent work by Fronsdal \cite{Fronsdal:1978vb} in their pioneering group theoretic study. They have established a most relevant theorem regarding the direct product of simplest (Di and Rac) representations of the conformal group, which were shown to decompose into an infinite sum of higher spin representations. This theorem represents the kinematical backbone of the more recent AdS/CFT correspondence where conformal fields through a large $N$ collective effect produce higher spin fields. The dipole picture was also identified in studies of high energy scattering in QCD \cite{Nikolaev:1993ke}. It has also appeared in studies of noncommutative field theory \cite{Rey:2001nq}.

The content of the present paper is as follows: in section two we describe the collective dipole as a two-body system with constraints and discuss gauge fixing to time-like or light-cone gauges. In section three we describe the system representing a higher spin particle in AdS. In section four we explain how the two systems are transformed into one another through a one-to-one map. This completes the demonstration that $d+1$ dimensional AdS spacetime and higher spins are generated in terms of the $d$ dimensional collective dipole. We end with conclusions in section five.

\section{The Collective Dipole}

The large $N$ quantum field theory of the $O(N)$ vector model
\begin{equation}
L=\int d^d x {1 \over 2}(\partial_\mu \phi^i)(\partial^\mu \phi^i)+V(\phi \cdot \phi), \qquad i=1,...,N
\end{equation}
represents a relatively simple field theory for critical phenomena and more recently as a model of AdS$_{d+1}$/CFT$_d$ correspondence. In three dimensions besides the trivial, free field theory UV fixed point one also has a nontrivial IR fixed point (for detailed studies see \cite{Lang:1990ni}). The AdS/CFT duality with Vasiliev's higher spin theory for both fixed points was understood in \cite{Klebanov:2002ja} and subsequent more recent work. In any dimension $d$ one can consider the free theory which nevertheless in the large $N$ limit exhibits a duality with a theory of higher spin in $d+1$ dimensions. The origin of higher spins and of the emergence of the extra radial AdS spatial dimension was given \cite{Das:2003vw} in terms of bilocal (collective) fields
\begin{equation}
\Psi(x_1^\mu,x_2^\nu)=\sum^N_{i=1}\phi^i(x_1^\mu)\phi^i(x_2^\nu),
\end{equation}
where $\mu,\nu=0,1,\cdots,d-1$ with the metric $(-,+,\cdots,+)$. These fields close a set of Schwinger-Dyson equations with an effective action that leads to a systematic $1/N$ expansion \cite{deMelloKoch:1996mj}. It was argued in \cite{Das:2003vw, Koch:2010cy} that this provides a bulk description of the AdS$_4$ dual higher-spin gravity (for the two conformal fixed points of the three dimensional field theory). This picture was sharpened in the time-like or null-plane quantization scheme, where the bilocal field involves a single time
\begin{equation}
\Psi(t,\vec{x}_1,\vec{x}_2)=\sum^N_{i=1}\phi^i(t,\vec{x}_1)\phi^i(t,\vec{x}_2).
\end{equation}
In this case a precise one-to-one map was formulated in \cite{Koch:2010cy} relating the light-cone higher spin field in AdS$_4$ and the collective bilocal field
\begin{eqnarray}
\Phi(x^-,x,z,\theta)&=&\int dp^+ dp^x dp^z e^{i (x^- p^++x p^x+z p^z)} \cr
&&\int dp_1^+ dp_2^+ dp_1 dp_2 \delta(p_1^+ + p_2^+ - p^+)\delta(p_1+p_2-p^x) \cr
&&\delta\Bigl(p_1 \sqrt{p_2^+ / p_1^+} -p_2 \sqrt{p_1^+ / p_2^+}-p^z\Bigr) \cr
&&\delta\bigl(2\arctan\sqrt{p_2^+ / p_1^+}-\theta\bigr) \tilde{\Psi}(p_1^+,p_2^+,p_1,p_2),
\label{LCmapping4}
\end{eqnarray}
where $\tilde{\Psi}(p_1^+,p_2^+,p_1,p_2)$ is the Fourier transform of the bilocal field $\Psi(x_1^-,x_2^-,x_1,x_2)$.

The physical basis of the correspondence can then be identified by a bi-particle system of a collective dipole which through a canonical transformation maps into the first quantization version of the higher spin system. In what follows, we would like to discuss and study this dipole construction in full detail. Our goal is to establish a first quantized or rather a world-sheet description of the AdS/CFT construction developed in \cite{Das:2003vw, Koch:2010cy}.

Let us start with a two-particle system in $d$-dimensional Minkowski spacetime with the action
\begin{equation}
S=\int d\tau_1 m \sqrt{\vert\dot{x}_1^2(\tau_1)\vert}+\int d\tau_2 m \sqrt{\vert\dot{x}_2^2(\tau_2)\vert}
\end{equation}
which leads to the constraints
\begin{eqnarray}
p_1^2+m^2&=&0, \label{onshell1} \\
p_2^2+m^2&=&0. \label{onshell2}
\end{eqnarray}
Switching to the center-of-mass variables
\begin{eqnarray}
P=p_1+p_2, &\qquad& X=\frac{1}{2}(x_1+x_2), \\
p=p_1-p_2, &\qquad& ~x=\frac{1}{2}(x_1-x_2),
\end{eqnarray}
the constraints (\ref{onshell1}-\ref{onshell2}) become
\begin{eqnarray} 
T_1&=&P^2+p^2+4m^2=0, \label{constraint1} \\
T_2&=&P\cdot p=0. \label{constraint2}
\end{eqnarray}
This system written in the above covariant form is described with two times. It therefore can potentially have problems with unitarity and appearance of ghosts as discussed in the investigations of \cite{Bars:1997bz}. In the present simple system one has the existence of a canonical gauge in which one can eliminate (gauge fix) the relative time and obtain a physical picture with a single time. This is analogous to the (second-quantized) collective field theory where one also had a covariant and a canonical, equal-time representation \cite{Jevicki:1979mb}.

Let us describe the details of such gauge fixing procedure, it was given some time ago \cite{Kamimura:1977dv} in connection with the investigation of Yukawa's bilocal field theory. One introduces the condition
\begin{equation}
T_3=P \cdot x=0. \label{constraint3}
\end{equation}
Then the constraints (\ref{constraint2}, \ref{constraint3}) become second class while (\ref{constraint1}) remains first class. If one considers the interacting problem with $m=m(x^2)$, then the above condition arises from the the Poisson commutation of (\ref{constraint1}) and (\ref{constraint2}). 

Next, taking $P_\mu$ to be time-like, we can explicitly solve the second class constraints and eliminate the relative time coordinate. First, one makes a canonical transformation
\begin{eqnarray}
P_\mu&=&P_\mu, \label{can1} \\
X^\mu&=&\tilde{X}^\mu+u^Lb^{\mu s}\pi_s-\frac{\pi_L}{P^2}b^{~\mu}_s u^s-u^r\pi_s b^{~\nu}_r \frac{\partial{b_\nu^{~s}}}{\partial{P_\mu}}+\frac{u^L\pi_L}{P^2}P^\mu, \\
p_\mu&=&\frac{P_\mu}{P^2}\pi_L+b_\mu^{~r} \pi_r, \\
x^\mu&=&P^\mu u^L+b^{~\mu}_r u^r, \label{can4}
\end{eqnarray}
with $r,s=1,...,d-1$ and $b_r^{~\mu}$ satisfying
\begin{eqnarray}
b_r^{~\mu} P_\mu&=&0, \label{bcon1} \\
b_{\mu r}b_s^{~\mu}&=&g_{rs}=(+,\cdots,+), \\
b_{~r}^{\mu}b_\nu^{~r}&=&g^\mu_{~\nu}-\frac{P^\mu P_\nu}{P^2}. \label{bcon3}
\end{eqnarray}
One can easily see that $u^L$, $\pi_L$ are the components parallel to $P_\mu$ while $u^r$, $\pi_r$ are normal to $P_\mu$.
Then the constraints (\ref{constraint2}) and (\ref{constraint3}) lead to $u^L=\pi_L=0$. Therefore the system can be described only using the center-of-mass coordinates ($\tilde{X}^\mu, P_\mu$) and the relative (spatial) coordinates ($\vec{u}, \vec{\pi}$). The canonical transformation (\ref{can1}-\ref{can4}) is simplified to be
\begin{eqnarray}
P_\mu&=&P_\mu, \label{cordtrans1} \\
X^\mu&=&\tilde{X}^\mu-u^r\pi_s b_r^{~\nu} \frac{\partial{b}_\nu^{~s}}{\partial{P_\mu}}, \\
p_\mu&=&b_\mu^{~r} \pi_r, \\
x^\mu&=&b_r^{~\mu} u^r. \label{cordtrans2}
\end{eqnarray}
For the massless case where $m=0$, the conformal generators of the bi-particle system are given by
\begin{eqnarray}
\hat{P}^\mu&=&p_1^\mu+p_2^\mu, \\
\hat{M}^{\mu\nu}&=&x_1^\mu p_1^\nu-x_1^\nu p_1^\mu+x_2^\mu p_2^\nu-x_2^\nu p_2^\mu, \\
\hat{D}&=&x_1^\mu p_{1\mu}+x_2^\mu p_{2\mu}, \\
\hat{K}^\mu&=&(x_1^\nu p_{1\nu})x_1^\mu-{1 \over 2}(x_1^\nu x_{1\nu})p_1^\mu+(x_2^\nu p_{2\nu})x_2^\mu-{1 \over 2}(x_2^\nu x_{2\nu})p_2^\mu,
\end{eqnarray}
where we have neglected the scaling constant term for simplicity. It is instructive to find the explicit form of the conformal generators. Choosing a solution satisfied by (\ref{bcon1}-\ref{bcon3}) as follows
\begin{equation}
b_{0r}=\frac{P_r}{\sqrt{\vert P^2 \vert}}, \qquad b_{ir}=\delta_{ir}-\frac{P_i P_r}{P^2+P^0\sqrt{\vert P^2 \vert}}, \qquad \mu=(0,i)
\end{equation}
one achieves a single time ($X^0=\tilde{X}^0=t$) formulation of the conformal generators
\begin{eqnarray}
\hat{P}^0&=&P^0=\sqrt{\vec{P}^2+\vec{\pi}^2}, \label{cogen1} \\
\hat{P}^i&=&P^i, \\
\hat{M}^{0i}&=&t P^i-\tilde{X}^i P^0+\frac{1}{\sqrt{\vert P^2 \vert}-P^0}(u^i P^s \pi_s-\pi^i P_r u^r), \\
\hat{M}^{ij}&=&\tilde{X}^i P^j-\tilde{X}^j P^i+u^i \pi^j-u^j \pi^i, \\
\hat{D}&=&-t P^0+\tilde{X}^i P_i+u^i \pi_i, \\
\hat{K}^0&=&-{1 \over 2}t^2 P^0+t(\tilde{X}^i P_i+u^i \pi_i)+\frac{1}{\sqrt{\vert P^2 \vert}-P^0}(\tilde{X}^i u_i P^s \pi_s-\tilde{X}^i \pi_i P_r u^r) \cr
&&-{1 \over 2}P^0[\tilde{X}^i \tilde{X}_i+u^i u_i]+\frac{2\sqrt{\vert P^2 \vert}-P^0}{2P^2(\sqrt{\vert P^2 \vert}-P^0)^2}(u^i P^s \pi_s-\pi^i P_r u^r)^2, \\
\hat{K}^i&=&{1 \over 2} t^2 P^i+t[-P^0 \tilde{X}^i+\frac{1}{\sqrt{\vert P^2 \vert}-P^0}(u^i P^s \pi_s-\pi^i P_r u^r)] \cr
&&+\tilde{X}^i [\tilde{X}^j P_j+u^j \pi_j]-\pi^i[\tilde{X}^j u_j+\frac{1}{P^2+P^0\sqrt{\vert P^2 \vert}}u^j u_j P^s \pi_s] \cr
&&+u^i[\tilde{X}^j \pi_j+\frac{1}{P^2+P^0\sqrt{\vert P^2 \vert}}(2 u^j \pi_j P^s \pi_s-\pi^j \pi_j P_r u^r)] \cr
&&-{1 \over 2}P^i\bigl[\tilde{X}^j \tilde{X}_j+u^j u_j+\frac{1}{P^2(\sqrt{\vert P^2 \vert}-P^0)^2}(u^i P^s \pi_s-\pi^i P_r u^r)^2\bigr]. \label{cogen2}
\end{eqnarray}
Now recall the canonical, equal-time $(x_1^0=x_2^0=t)$ collective field version of the bi-particle system, where the conformal generators are given by
\begin{eqnarray}
\hat{P}^0&=&p_1^0+p_2^0=\sqrt{\vec{p}_1{}^2}+\sqrt{\vec{p}_2{}^2}, \label{bigen1} \\
\hat{P}^i&=&p_1^i+p_2^i, \\
\hat{M}^{0i}&=&t (p_1^i+p_2^i)-x_1^i p_1^0-x_2^i p_2^0, \\
\hat{M}^{ij}&=&x_1^i p_1^j-x_1^j p_1^i+x_2^i p_2^j-x_2^j p_2^i, \\
\hat{D}&=&-t(p_1^0+p_2^0)+x_1^i p_1^i+x_2^i p_2^i, \\
\hat{K}^0&=&-{1 \over 2}t^2(p_1^0+p_2^0)+t(x_1^i p_1^i+x_2^i p_2^i) \cr
&&-{1 \over 2}x_1^i x_1^i p_1^0-{1 \over 2}x_2^i x_2^i p_2^0, \\
\hat{K}^i&=&{1 \over 2}t^2(p_1^i+p_2^i)-t(x_1^i p_1^0+x_2^i p_2^0) \cr
&&+x_1^j p_1^j x_1^i+x_2^j p_2^j x_2^i-{1 \over 2}x_1^j x_1^j p_1^i-{1 \over 2} x_2^j x_2^j p_2^i. \label{bigen2}
\end{eqnarray}
There is a simple canonical transformation between the phase space ($\tilde{X}^i, P^i; u^i, \pi^i$) and the bi-particle phase pace ($x_1^i, p_1^i; x_2^i, p_2^i$), which transforms the generators (\ref{cogen1}-\ref{cogen2}) to (\ref{bigen1}-\ref{bigen2}). 
It is given by
\begin{eqnarray}
P^i&=&p_1^i+p_2^i, \\
\tilde{X}^i&=&{x_1^i p_1^0+x_2^i p_2^0 \over p_1^0+p_2^0}+{1 \over P^0(P^2+P^0\sqrt{\vert P^2 \vert})} \cr
&&\times[(x_1^i-x_2^i)(p_1^j p_2^0-p_2^j p_1^0)(p_1^j+p_2^j) \cr
&&-(p_1^i p_2^0-p_2^i p_1^0)(x_1^j-x_2^j)(p_1^j+p_2^j)], \\
\pi^i&=&-\frac{\sqrt{\vert P^2 \vert}-2p_2^0}{\sqrt{\vert P^2 \vert}-P^0} p_1^i+\frac{\sqrt{\vert P^2 \vert}-2p_1^0}{\sqrt{\vert P^2 \vert}-P^0} p_2^i, \\
u^i&=&-{1 \over 2}(x_1^i-x_2^i)-{p_1^0-p_2^0 \over (P^0)^2P^2}(x_1^j-x_2^j)(p_1^j+p_2^j)(p_1^i p_2^0-p_2^i p_1^0) \cr
&&+{2 p_1^0 p_2^0 \over (P^0)^2(P^2+P^0\sqrt{\vert P^2 \vert})}(x_1^j-x_2^j)(p_1^j+p_2^j)(p_1^i+p_2^i).
\end{eqnarray}

We have in the above described the canonical structure of the composite, two particle ``collective'' dipole system. It was constructed to describe the singlet subspace of the vector model CFT. Since the CFT has two collective field representations (one covariant with an associated action and another equal-time with an effective Hamiltonian) it was important to demonstrate the existence of a single time gauge. We have also seen that in this gauge the dipole system exhibits an identical canonical structure to the collective field theory one. This structure is characterized by an additive contribution to the symmetry generators which we established. In the next section we will review the field theory of higher spins in AdS and describe its first quantized description given by particles with spin moving in AdS spacetime. We will then establish in section four that the $d$-dimensional dipole system is canonically related to the $d+1$ dimensional AdS system through a one-to-one transformation.

\section{Higher Spin Theory in AdS}

We now switch to a discussion of higher spin theory in $AdS_{d+1}$ spacetime. From the field theoretic description of this theory we will deduce a first-quantized AdS particle system (with spin). We will then demonstrate in section four that the AdS spin particle system emerges through a canonical change of variables from the $d$-dimensional collective dipole system.

\subsection{Higher Spin Fields}

There are two formalisms for describing higher spin fields, one being the frame-like formulation in terms of generalized vielbeins and spin connections, the other the metric-like formulation due to Fronsdal \cite{Fronsdal:1978rb}, which employs higher tensor fields with arbitrary rank and symmetry properties. Here, we will use the second formulation. One also has interesting alternative approaches studied in \cite{Francia:2002aa}.

A spin $s$ field is represented by a symmetric and double traceless tensor of rank $s$: $h_{\mu_1...\mu_s}(x^\mu)$, which obeys the equations of motion \cite{Mikhailov:2002bp}
\begin{eqnarray}
\nabla_\rho \nabla^\rho h_{\mu_1...\mu_s}-s\nabla_\rho \nabla_{\mu_1}h^\rho_{~\mu_2...\mu_s}+\frac{1}{2}s(s-1)\nabla_{\mu_1} \nabla_{\mu_2}h^\rho_{~\rho \mu_3...\mu_s} \cr
+2(s-1)(s+d-2)h_{\mu_1...\mu_s}=0.
\label{HSeom}
\end{eqnarray}
The gauge transformation is given by
\begin{equation}
\delta_{\Lambda}h^{\mu_1...\mu_s}=\nabla^{\mu_1}\Lambda^{\mu_2...\mu_s}, \qquad g_{\mu_2 \mu_3} \Lambda^{\mu_2...\mu_s}=0.
\end{equation}
A covariant gauge can be specified with the gauge conditions
\begin{equation}
\nabla^\rho h_{\rho \mu_2...\mu_s}=0, \qquad g^{\rho \sigma}h_{\rho \sigma \mu_3...\mu_s}=0.
\end{equation}
Then the equation of motion (\ref{HSeom}) reduces to
\begin{equation}
(\square+m^2)h_{\mu_1...\mu_s}=0,
\end{equation}
with the effective mass $m^2=s^2+(d-5)s-2(d-2)$.

It is useful to embed the $d+1$ dimensional AdS spacetime $x^\mu$ into $d+2$ dimensional hyperboloid $x^\alpha$ with the metric $(-,+,...,+,-)$. The higher spin field $h_{\mu_1...\mu_s}(x^\mu)$ is related to the after-embedding higher spin field $k_{\alpha_1...\alpha_s}(x^\alpha)$ by
\begin{equation}
k_{\alpha_1 ... \alpha_s}(x^\alpha)=x^{~\mu_1}_{\alpha_1} \cdots x^{~\mu_s}_{\alpha_s} h_{\mu_1 ... \mu_s}(x^\mu),
\end{equation}
where $x^{~\mu}_{\alpha}=\partial x^\mu / \partial x^\alpha$. Introducing an internal set of coordinates $y^\alpha$ spacetime, one forms the field with all spins
\begin{equation}
K(x^\alpha,y^\alpha) \equiv \sum_s k_{\alpha_1...\alpha_s}(x^\alpha)y^{\alpha_1} \cdots y^{\alpha_s}.
\end{equation}
In this notation the constraints implied by embedding, the covariant gauge conditions as well as the equations of motion become the following system of equations \cite{Fronsdal:1978vb}
\begin{eqnarray}
&&\partial_x^2 K(x,y)=0, \label{froncons1} \\
&&\partial_y^2 K(x,y)=0, \\
&&\partial_x \cdot \partial_y K(x,y)=0, \\
&&(x \cdot \partial_x+y \cdot \partial_y+1)K(x,y)=0, \\
&&x \cdot \partial_y K(x,y)=0. \label{froncons5}
\end{eqnarray}
It is easy to check that the constraints (\ref{froncons1}-\ref{froncons5}) are all first-class constraints. We should also point out that $\Phi(x^-,x,z,\theta)$ in (\ref{LCmapping4}) is the light-cone form of $K(x^\alpha,y^\alpha)$ in AdS$_4$.

In the above representation one has an asymmetry between the spacetime coordinates $x$ and the internal spin coordinates $y$ due to (\ref{froncons5}). One can through a series of canonical transformations achieve a totally symmetric description. The transformation was found by Fronsdal in \cite{Fronsdal:1978vb}, which takes the form
\begin{equation}
\Phi(p,q)=(FK)(x,y)
\end{equation}
where $p=(x+y)/2$, $q=(x-y)/2$ and the kernel for a particular spin $s$ is given by
\begin{equation}
F_s=\sum_k (4^k k!)^{-1} (y \cdot \partial_x)^{2k}/(\hat{n}+1)(\hat{n}+2)\cdots(\hat{n}+k)
\label{Fronsmap}
\end{equation}
with $\hat{n}=y \cdot \partial_y$. After the mapping (\ref{Fronsmap}), as well as a Fourier transformation
\begin{equation}
\Phi(u,v)=\int dp dq~ e^{i p \cdot u+i q \cdot v} \Phi(p,q),
\end{equation}
one finds the symmetric version 
\begin{eqnarray} 
&&(u \cdot \partial_u+1/2)\Phi(u,v)=0, \label{fronsym1} \\
&&(v \cdot \partial_v+1/2)\Phi(u,v)=0, \label{fronsym2} \\
&&u^2=0, \label{fronsym3} \\
&&v^2=0, \label{fronsym4} \\
&&u \cdot v=0. \label{fronsym5}
\end{eqnarray} 

Next we show that it is possible to reduce the system by solving the first four constraints (\ref{fronsym1}-\ref{fronsym4}) which are decoupled into two sets of constraints involving separately $u$ and $v$. Parameterizing the cone $u^2=0$ as
\begin{equation}
u_0=U\sin t, \qquad u_{d+1}=U\cos t, \qquad \vec{u}=U\hat{u},
\end{equation}
with $\hat{u}^2=1$, we find the constraint (\ref{fronsym1}) becomes $\partial / \partial U+1/2$. Consequently the dependence on the variable $U$ can be factored out
\begin{equation}
\phi(u)=U^{-1/2}\phi(t,\hat{u}),
\end{equation}
and the remaining degrees of freedom are the coordinate $(t,\hat{u})$ (and its conjugates). Similarly, this reduction works for the $v$ system. Therefore, by solving the first four constraints, we reduced the bilocal field $\Phi(u,v)$ with $2(d+2)$ variables to $2d$ variables. This agrees precisely with the bilocal collective field $\Phi(x_1^\mu,x_2^\mu)$ in $d$ dimensions. However, in this formulation, we need to interpret (\ref{fronsym5}) as the equation of motion, which does not take the form of the collective equation of motion \cite{Das:2003vw}. In order to make contact with the collective field equation, one can trade in (\ref{fronsym5}) with a new constraint
\begin{equation}
\partial_u^2 \partial_v^2 \Phi(u,v)=0,
\end{equation}
which does not commute with (\ref{fronsym5}). As shown in \cite{Das:2003vw}, this is the equation of motion for the collective field after a field redefinition. This shows that the bilocal collective field theory of \cite{Das:2003vw} corresponds to another gauge choice when compared with the Fronsdal's covariant gauge higher spin theory.

\subsection{Higher Spin Particles in AdS$_4$}

To describe particles in AdS with spin, one uses the spacetime coordinate $x$ and an internal spin coordinate $y$. For simplicity, we will mainly discuss the AdS$_4$ case (only in this subsection), which corresponds to the isometry group $SO(2,3)$. The system requires constraints expressing strong conservation of the phase space counterparts of the second- and fourth-order Casimir operators of $so(2,3)$. We have the generators
\begin{equation}
J_{AB}=x_A p^x_B-x_B p^x_A+y_A p^y_B-y_B p^y_A
\end{equation}
where $x^A$ and $y^A$ represent two separate objects and $A,B=0,1,2,3,5$ with the metric $\eta_{AB}={\rm diag}(-,+,+,+,-)$. The second- and fourth-order Casimir operators are given by
\begin{eqnarray}
\Omega_1&=&{1 \over 2}J_{AB} J^{AB} \cr
&=&x^2 p_x^2-(x \cdot p_x)^2+y^2 p_y^2-(y \cdot p_y)^2 \cr
&&+2(x \cdot y)(p_x \cdot p_y)-2(x \cdot p_y)(y \cdot p_x), \\
\Omega_2&=&{1 \over 4}J_{AB} J^B {}_C J^C {}_D J^{DA}-{1 \over 2}\Bigl({1 \over 2}J_{AB} J^{AB}\Bigr)^2 \cr
&=&x^2(p_y^2(yp_x)^2+p_x^2(yp_y)^2-2(p_x p_y)(y p_x)(y p_y)) \cr
&&+y^2(p_y^2(xp_x)^2+p_x^2(xp_y)^2-2(p_x p_y)(x p_x)(x p_y)) \cr
&&+x^2 y^2((p_x p_y)^2-p_x^2 p_y^2)+(x y)^2(p_x^2 p_y^2-(p_x p_y)^2) \cr
&&-(x p_y)^2 (y p_x)^2-(x p_x)^2 (y p_y)^2+2(x p_x)(x p_y)(y p_x)(y p_y) \cr
&&+2(p_x p_y)(x p_x)(xy)(yp_y)+2(p_x p_y)(x p_y)(x y)(y p_x) \cr
&&-2 p_x^2(x p_y)(x y)(y p_y)-2 p_y^2(x p_x)(x y)(y p_x).
\end{eqnarray}
They are constrained to
\begin{eqnarray}
\Omega_1+E_0^2+s^2&=&0, \label{hsparcon1} \\
\Omega_2+E_0^2 s^2&=&0. \label{hsparcon2}
\end{eqnarray}
One solution to the constraints (\ref{hsparcon1}-\ref{hsparcon2}) leads to
\begin{eqnarray}
x \cdot p_x&=&-E_0, \label{cvgauge1} \\
x \cdot p_y&=&0, \label{cvgauge2} \\
y \cdot p_y&=&s, \label{cvgauge3} \\
p_x^2&=&0, \label{cvgauge4} \\
p_y^2&=&0, \label{cvgauge5} \\
p_x \cdot p_y&=&0.\label{cvgauge6} 
\end{eqnarray}
The massless higher spin particle corresponds to the special case $E_0=s+1$. These constraints are seen to agree with Fronsdal's covariant formulation of higher spin theory (\ref{froncons1}-\ref{froncons5}). Another canonical representation of the higher-spin particle system solving the constraints (\ref{hsparcon1}-\ref{hsparcon2}) was given in \cite{Kuzenko:1995aq}
\begin{eqnarray}
x^2+r^2 &=& 0, \label{surface1} \\
x \cdot p_x &=& 0, \label{surface2} \\
x \cdot y &=& 0, \label{surface3} \\
x \cdot p_y &=& 0, \label{surface4} \\
y \cdot p_y &=& 0, \\
p_y^2 &=& 0, \\
p_x^2 &=& \frac{E_0^2+s^2}{r^2}, \\
p_x \cdot p_y &=& \Bigl(\frac{E_0^2s^2}{r^2y^2}\Bigr)^{1/2},      
\end{eqnarray}
where $r$ is the radius of the AdS spacetime and (\ref{surface1}, \ref{surface3}) are gauge conditions for the first-class constraints (\ref{surface2}, \ref{surface4}) respectively.

\section{AdS$_{d+1}$ from $d$-dimensional dipole}

We now come to the main part of our construction. We will show (in the framework of the light-cone gauge) that $d$-dimensional relativistic bi-particle system of section two can be mapped into the higher spin AdS$_{d+1}$ particle system. This map will be accomplished by an explicit canonical transformation between the respective phase space variables. In the process we will be able to map the collective field version of the conformal generators to the generators that can be constructed in AdS spacetime.

For specifying the light-cone gauge of higher spin theory in AdS$_{d+1}$, one starts following \cite{Metsaev:1999ui} with the gauge invariant description with the AdS and internal coordinates denoted by $(x_{\hat{\mu}},p^{\hat{\mu}},\bar{\alpha}_{\hat{\mu}},\alpha^{\hat{\mu}})$, $\hat{\mu}=0,1,2,..,d$. One can parametrize the AdS$_{d+1}$ space with the Poincar\'e coordinates
\begin{equation}
dx_{\hat{\mu}}dx^{\hat{\mu}}=\frac{1}{z^2}(-dt^2+dx_i^2+dz^2+dx_d^2), \qquad i=1,...,d-2.
\end{equation}
The light-cone variables and transverse coordinates are denoted as
\begin{equation} 
x^\pm =\frac{1}{\sqrt{2}}(x^d \pm x^0), \qquad x^I=(x^i,z).
\end{equation}
The light-cone gauge \cite{Metsaev:1999ui} is now fully specified by the conditions
\begin{eqnarray}
&&\bar{\alpha}^+=0, \label{lcgauge1} \\
&&\alpha^I \bar{\alpha}^I=s, \label{lcgauge2} \\
&&\bar{\alpha}^I \bar{\alpha}^I=0, \label{lcgauge3} \\
&&\bar{\alpha}^-=-\frac{p^I}{p^+}\bar{\alpha}^I+\frac{s+d-1}{p^+}\bar{\alpha}^z-\frac{2(p^+-\alpha^+\bar{\alpha}^z)}{p^+(p^+-2\alpha^+\bar{\alpha}^z)}\bar{\alpha}^z, \label{lcgauge4} \\
&&(p^{\hat{\mu}}-\alpha^{\hat{\mu}}\bar{\alpha}^z)^2-(2\alpha^z-\alpha^{\hat{\mu}} p^{\hat{\mu}}-\alpha^z\alpha^{\hat{\mu}} \bar{\alpha}^{\hat{\mu}}+\alpha^2\bar{\alpha}^z)\frac{2(p^+-\alpha^+\bar{\alpha}^z)}{p^+(p^+-2\alpha^+\bar{\alpha}^z)}\bar{\alpha}^z\cr 
&&-d(p^z-\alpha^z \bar{\alpha}^z)-s^2+(4-d)s+2d-4=0. \label{lcgauge5}
\end{eqnarray}
Here (\ref{lcgauge1}) represents the light-cone gauge condition, and the constraints (\ref{lcgauge2}, \ref{lcgauge3}, \ref{lcgauge5}) are analogous to (\ref{cvgauge3}, \ref{cvgauge5}) and (\ref{cvgauge4}) in our particle description respectively. From the Lorentz condition (\ref{cvgauge6}), one can solve for $\bar{\alpha}^-$ (\ref{lcgauge4}). For more detailed studies of light-cone higher spin theory in AdS spacetime the reader should consult \cite{Metsaev:1999ui}.

Our construction of the canonical relationship between the two sets of variables will come from the comparison of two different
 representations of the generators of the conformal group: one corresponding to the $d$-dimensional dipole and the other to the $d+1$ dimensional higher-spin AdS particle. For this we first recall the light-cone form of generators in AdS given by  \cite{Metsaev:1999ui}
\begin{eqnarray} 
\hat{P}^-&=&-\frac{p_I^2}{2p^+}-\frac{1}{2z^2 p^+}(\frac{1}{2}m_{ij}^2-\frac{1}{4}(d-3)(d-5)), \label{adsd1} \\
\hat{P}^+&=&p^+, \label{adsd2} \\
\hat{P}^i&=&p^i, \label{adsd3} \\
\hat{J}^{+-}&=&t\hat{P}^--x^- p^+, \label{adsd4}\\
\hat{J}^{+i}&=&t p^i-x^i p^+,\label{adsd5} \\
\hat{J}^{-i}&=&x^- p^i-x^i \hat{P}^-+m^{iJ}\frac{p^J}{p^+}-\frac{1}{2z p^+}\{m^{zj},m^{ji}\},\label{adsd6}  \\
\hat{J}^{ij}&=&x^i p^j-x^j p^i+m^{ij}, \label{adsd7} \\
\hat{D}&=&t \hat{P}^-+x^- p^++x^I p^I+\frac{d-1}{2}, \label{adsd8} \\
\hat{K}^-&=&-\frac{1}{2}x_I^2 \hat{P}^- +x^- (x^- p^++x^I p^I+\frac{d-1}{2}) \cr
&&+\frac{1}{p^+}x^I p^J m^{IJ}-\frac{x^I}{2z p^+}\{m^{zJ},m^{JI}\}, \label{adsd9} \\
\hat{K}^+&=&t^2 \hat{P}^-+t (x^I p^I+\frac{d-1}{2})-\frac{1}{2}x_I^2 p^+, \label{adsd10} \\
\hat{K}^i&=&t(x^i \hat{P}^--x^- p^i-m^{iJ}\frac{p^J}{p^+}+\frac{1}{2z p^+}\{m^{zj},m^{ji}\}) \cr
&&-\frac{1}{2}x_J^2 p^i+x^i (x^- p^++x^I p^I+\frac{d-1}{2})+m^{iI}x^I. \label{adsd11}
\end{eqnarray}
These generators are to be compared with the bilocal CFT$_d$ transformations. In the light-cone gauge ($x_1^+=x_2^+=t$), one has the bilocal collective field
\begin{equation}
\Psi(x_1^\mu,x_2^\nu) \mapsto \Psi(t;x_1^-,x_1^i;x_2^-,x_2^j).
\end{equation}
The conformal generators take the form
\begin{eqnarray}
\hat{P}^-&=&p_1^-+p_2^-=-\Bigl({p_1^i p_1^i \over 2p_1^+}+{p_2^i p_2^i\over 2p_2^+}\Bigr), \label{cftd1} \\
\hat{P}^+&=&p_1^++p_2^+, \label{cftd2} \\
\hat{P}^i&=&p_1^i+p_2^i, \label{cftd3} \\
\hat{J}^{+-}&=&t\hat{P}^--x_1^- p_1^+-x_2^- p_2^+, \label{cftd4} \\
\hat{J}^{+i}&=&t\hat{P}^i-x_1^i p_1^+ -x_2^i p_2^+, \label{cftd5} \\
\hat{J}^{-i}&=&x_1^- p_1^i+x_2^- p^i_2+x_1^i{p^j_1 p_1^j\over 2p_1^+}+x_2^i{p^j_2 p_2^j \over 2p_2^+}, \label{cftd6} \\
\hat{J}^{ij}&=&x_1^ip_1^j-x_1^jp_1^i+x_2^ip_2^j-x_2^jp_2^i, \label{cftd7} \\
\hat{D}&=&t\hat{P}^-+x_1^- p_1^+ + x_2^- p_2^+ +x_1^i p_1^i+x_2^i p_2^i+ 2d_\phi, \label{cftd8} \\
\hat{K}^-&=&x_1^i x_1^i{p^j_1 p_1^j \over 4p_1^+}+x_2^i x_2^i{p^j_2 p_2^j \over 4p_2^+}+x_1^-(x_1^- p_1^++x_1^i p_1^i+d_\phi) \cr
&&+x_2^-(x_2^- p_2^++x_2^i p_2^i+d_\phi), \label{cftd9} \\
\hat{K}^+&=&t^2 \hat{P}^-+t(x_1^i p_1^i+x_2^i p_2^i+2d_\phi)-{1 \over 2}x_1^i x_1^i p_1^+ -{1 \over 2}x_2^i x_2^i p_2^+, \label{cftd10} \\
\hat{K}^i&=&-t \Bigl(x_1^i{p^j_1 p_1^j \over 2p_1^+}+x_2^i{p^j_2 p_2^j \over 2p_2^+}+x_1^- p^i_1+x_2^- p^i_2\Bigr)-{1 \over 2}x_1^j x_1^j p_1^i-{1 \over 2} x_2^j x_2^j p_2^i \cr
&&+x_1^i(x_1^- p_1^++x_1^j p_1^j+d_\phi)+x_2^i(x_2^- p_2^++x_2^j p_2^j+d_\phi). \label{cftd11}
\end{eqnarray}

For simplicity, we will neglect the scale dimension terms on both sides, which can be added at the quantum level. Furthermore, the Poisson bracket $\{m^{zj},m^{ji}\}$ and $\{m^{zJ},m^{JI}\}$ should be simplified as $2m^{zj}m^{ji}$ and $2m^{zJ}m^{JI}$ respectively. The phase space on the two sides are ($x^-,x^i,z,\theta^{IJ};p^+,p^i,p^z,m^{IJ}$) and ($x_1^-,x_1^i,x_2^-,x_2^i;p_1^+,p_1^i,p_2^+,p_2^i$). The canonical transformation is found by comparing the higher-spin generators (\ref{adsd1}-\ref{adsd11}) with the collective dipole generators (\ref{cftd1}-\ref{cftd11}). The AdS coordinates and conjugate momenta are given by
\begin{eqnarray}
x^-&=&\frac{x_1^-p_1^++x_2^-p_2^+}{p_1^++p_2^+}, \\
p^+&=&p_1^++p_2^+, \\
x^i&=&\frac{x_1^i p_1^++x_2^i p_2^+}{p_1^++p_2^+}, \\
p^i&=&p_1^i+p_2^i, \\
z&=&\frac{\sqrt{p_1^+ p_2^+}}{p_1^++p_2^+}\sqrt{(x_1^i-x_2^i)^2}, \\
p^z&=&\frac{x_1^j-x_2^j}{\sqrt{(x_1^i-x_2^i)^2}}\Bigl(p_1^j\sqrt{p_2^+ \over p_1^+}-p_2^j\sqrt{p_1^+ \over p_2^+}\Bigr),
\end{eqnarray}
and the angular momenta are found to be
\begin{eqnarray}
m^{ij}&=&\frac{1}{p_1^++p_2^+}[(x_1^i-x_2^i)(p_1^jp_2^+-p_2^jp_1^+)-(x_1^j-x_2^j)(p_1^ip_2^+-p_2^ip_1^+)], \\
m^{iz}&=&\frac{x_1^i-x_2^i}{\sqrt{(x_1^j-x_2^j)^2}}\Bigl[\sqrt{p_1^+ p_2^+}(x_1^--x_2^-)+\frac{\bigl((p_1^+)^2 p_2^j+(p_2^+)^2 p_1^j)(x_1^j-x_2^j)}{(p_1^++p_2^+)\sqrt{p_1^+ p_2^+}}\Bigr] \cr
&+&\frac{1}{2}\frac{p_1^+-p_2^+}{p_1^++p_2^+}\sqrt{(x_1^j-x_2^j)^2}\Bigl(p_1^i\sqrt{p_2^+ \over p_1^+}-p_2^i\sqrt{p_1^+ \over p_2^+}\Bigr).
\end{eqnarray}
This is a canonical transformation as one can verify the Poisson brackets
\begin{eqnarray}
&&\{x^-,p^+\}=1, \qquad \{x^i,p^j\}=\delta^{ij}, \qquad \{z,p^z\}=1, \\
&&\{m^{IJ},m^{KL}\}=\delta^{JK}m^{IL}+\delta^{IL}m^{JK}-\delta^{JL}m^{IK}-\delta^{IK}m^{JL},
\end{eqnarray}
and others vanish.

In summary, we have in the above established a one-to-one map between the phase space coordinates of the collective dipole and the phase space of the higher spin AdS particle. This map can be seen as a concrete realization of the aforementioned theorem of Flato and Fronsdal \cite{Flato:1978qz} regarding the tensor product of Di-Rac representations of the conformal group. The theorem has been generalized to any dimension in \cite{Vasiliev:2004cm} as is our explicit construction.

This map generalizes the earlier construction established in $d=3$ to any dimension, which provides a simple explicit model of the AdS/CFT correspondence. In the gauge used (light-cone gauge) it reconstructs the AdS theory in the bulk. Issues of locality in the AdS spacetime have been studied recently in \cite{Kabat:2011rz}. This construction demonstrates how a non-local (bi-particle space) is transformed into the local AdS spacetime with higher spins.

\section{Conclusion}

We have in present paper described the collective dipole picture of the AdS/CFT correspondence. This picture was extracted from the bilocal field representation of a conformally invariant $O(N)$ vector model. These fields which fully describe the $O(N)$ singlet sector of the theory were seen to contain the full interacting bulk AdS theory with higher spins. A first quantized  description represents a bi-particle system which we called the collective dipole. We have studied the structure of constraints and gauge fixing of the dipole system. This issue itself is nontrivial as we are dealing with a fully relativistic system with two times. Consequently various issues related to unitarity and absence of ghosts have to be addressed. We have following earlier work discussed in detail the issue of gauge fixing to a physical time-like or light-cone gauge. Using a gauge condition which leads to elimination of the relative time, we have exhibited the existence of a unitary, ghost-free representation of the dipole system. This gauge also establishes contact with the equal-time Hamiltonian of bilocal field theory.

Using the light-cone frame we have then demonstrated the correspondence with the higher-spin particle in AdS spacetime. This correspondence is constructed in terms of an explicit one-to-one canonical map relating the $d$-dimensional collective dipole with the $d+1$ dimensional higher spin particle in AdS. The map gives an explicit reconstruction of the extra (radial) AdS space dimension and of the infinite sequence of higher spin states. As such it represents likely the simplest system where the AdS/CFT correspondence is established in the bulk.

For higher spin theory the relevance of the dipole picture lies in the following. It provides a first quantized world-sheet description of the theory and also has the promise to lead to a BRST quantization of the the system. The BRST approach was extremely relevant in the case of string theory \cite{Witten:1985cc} (and of course gauge theory \cite{Pashnev:1998ti}) but even though there have been various attempts there is not as yet a complete BRST description of Vasiliev's higher spin gauge theory. We plan to address this question in future publications.

\ack

We would like to thank J. Avan, C. I. Tan and especially S. Das for interesting discussions. Two of us (AJ and KJ) would like to thank the Simons Center for Geometry and Physics at Stony Brook for their hospitality during the workshop on Higher Spin Theories and Holography. This work is supported by the Department of Energy under contract DE-FG02-91ER40688.

\end{document}